# Enabling Autonomous Teams and Continuous Deployment at Scale[1]

*Torgeir Dingsøyr, Magne Jørgensen, Frode Odde Carlsen, Lena Carlström, Jens Engelsrud, Kine Hansvold, Mari Heibø-Bagheri, Kjetil Røe, Karl Ove Vika Sørensen*

*Abstract: In this article, we give advice on transitioning to a more agile delivery model for large-scale agile development projects based on experience from the Parental Benefit Project of the Norwegian Labour and Welfare Administration. The project modernized a central part of the organization's IT portfolio and included up to ten development teams working in parallel. The project successfully changed from using a delivery model which combined traditional project management elements and agile methods to a more agile delivery model with autonomous teams and continuous deployment. This transition was completed in tandem with the project execution. We identify key lessons learned which will be useful for other organizations considering similar changes and report how the new delivery model reduced risk and opened up a range of new possibilities for delivering the benefits of digitalization.*

As part of digitalization [1], many organizations, in both the private and public sector, are undergoing or planning an "agile transformation" with goals such as to delivering products with increased value and enabling the more frequent deployment of product improvements. An important part of this transition is frequently "scaling-up" the use of agile processes from team to project, program, and/or organizational level [2-5].

Several studies have been conducted on the transition to more agile processes and extension of agile processes to larger parts of the organization. Lessons learned from a successful transformation at the telecom company Ericsson include that the transition should be supported by using an experimental approach, stepwise implementation, specialized teams, and provision of sufficient support for teams to learn the agile processes [6]. Studies reporting on challenges in

---

such transitions find that agile processes may be difficult to implement, that requirements engineering may be more challenging, and that there may, in general, be resistance to change [4, 7]. However, we are not aware of any other study reporting on experience of managing this type of transition during the execution of a large project, with a fixed budget and planned date for delivery. This type of transition is not without peril, as we know there is a high risk for organizational change programs to fail [8].

**The parental benefit project**
In 2016, the Norwegian Labour and Welfare Administration ("Welfare Administration") initiated a large and highly publizised three-year project with an actual cost of around 75 million USD. The project aimed at developing software products which would process 100,000 applications for parental benefit annually, replacing a legacy solution. The development started in October 2016, and the software products were moved from project to product development organization in June 2019. At its peak, the development involved almost 130 people, and for a long period, ten development teams were working in parallel [9].

The first phase of the project used a delivery model with traditional project management elements, such as phases and subprojects, combined with agile process elements, such as iterative development and involvement of the business, in regular reviews [10]. In this phase, the requirements were described through epics, which were broken down into user stories and detailed by domain experts. The user stories were then handed over to the development teams (Figure 1, top). The teams used Scrum and Extreme Programming (XP) practices with three-week iterations. The solutions delivered in the first phase were deployed in two large releases in a complex release process.

Why change the delivery model? There was both internal and external pressure to change. Internally, many perceived the initial delivery model of the project to be a barrier to efficient work and argued for a more agile approach. The organization of the project led to many handovers between business and development subprojects, and, although a number of arenas were established, there were challenges in the coordination of these two subprojects. The delivery model used in the first phase of the project had few and very large releases, which meant long delays in feedback on the actual product from stakeholders. Furthermore, it was not clear how to use the delivery model to support operations and maintenance of the product in use while, in parallel, developing new features on related products. This was a high risk in the project. At the same time, the Welfare Administration had started work to change the general delivery model in favor of more agile approaches [11].

To face these challenges, the project manager established a task force to give advice on changing to a new, more agile, delivery model. The advice from the task force was central to the radical changes in the last phase of the project.

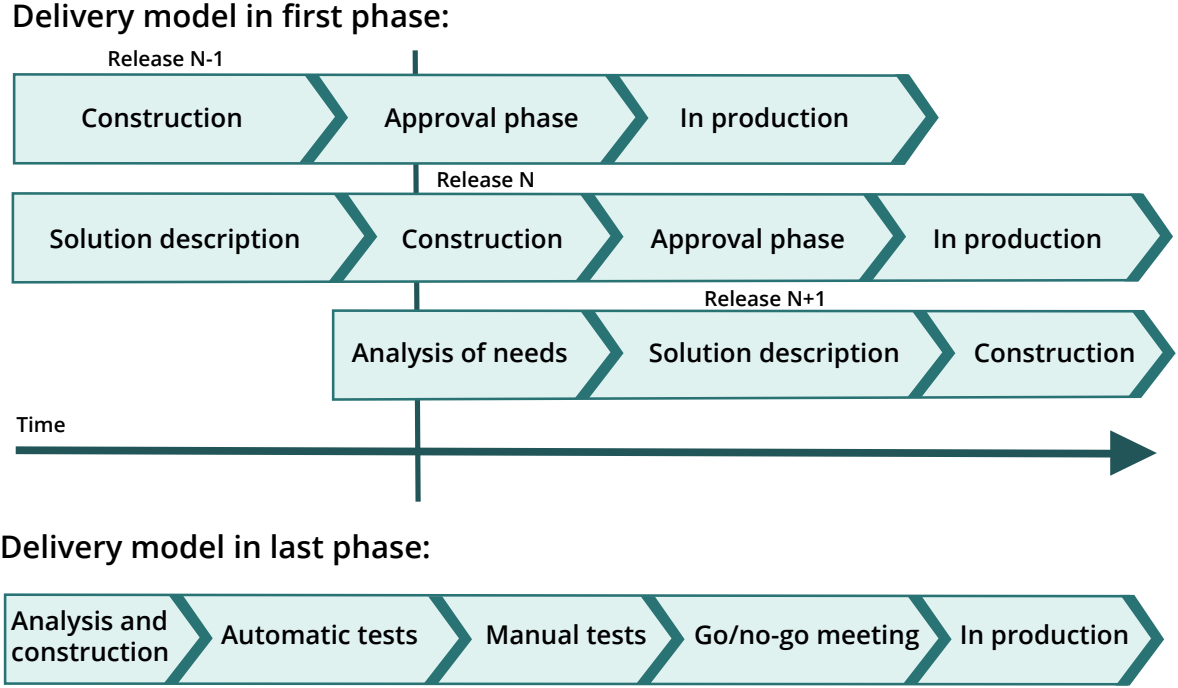

Figure 1: Delivery models in the first and last phases of the project.

In the last phase, the project changed the processes to a flow-based model with continuous deployment (Figure 1, bottom) [12]. Hence, both team and organizational design were changed, as were the technical platform and architecture, while the project milestones and value realization were retained. The organizational structure in the project changed from separating teams by role and phase to combining roles from business and development into cross-functional autonomous teams, separated by subdomain. The ten teams had, on average, 12 members and no formal team leadership. Each team had its own product owner.

The project was completed successfully, with all planned benefits realized on time and on cost, and received the annual prize for digital projects in Norway in 2019 for "making the user journey modern and digital for new and future parents." Users found that the time used to process applications was reduced from months to seconds. An example key objective was the degree of degree of self-service on applications which had a target of 80% but was 99.8% in the spring of 2019. The changes were also perceived as successful internally, where informants reported a "*much tighter dialogue*" between business and development and that "*it [the process change] strengthens the developers' understanding of the domain, the product owners' understanding of the technology, you save a lot of time and get more work done.*"

In the following, we describe what was experienced during the transition leading to the successful delivery of the project. We hope that the lessons learned from the transition will be valuable for other private and public organizations seeking to exploit the potential of more agile processes in ongoing large-scale digitization work.

**Start with a pilot team and a carefully selected delivery**
The project established a pilot team in a non-critical area to act as a frontrunner for new practices, gain experience, and learn about how to implement the new delivery model. The selection of the service subject to the pilot work was based on the five criteria shown in Figure 2. The pilot work was conducted in an autonomous, cross-functional team using a Kanban [13] approach with continuous deployments.

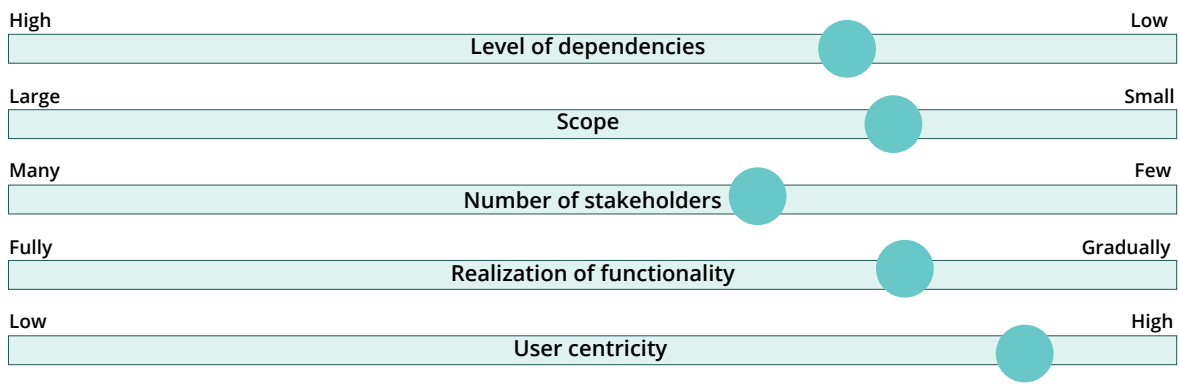

*Figure 2: Evaluation of self-service as pilot area.*

The use of the pilot team enabled the organization to gain important experience about the new delivery model and created momentum for the change. Importantly, this was done without severe consequences in the event of failure, as there was an existing solution that could act as a fallback. The pilot gave the team considerable freedom and leverage to experiment, potentially fail, and learn from experience.

One benefit of the pilot work was that the introduced processes were easy to adjust, based on what the pilot team and its surroundings had learned. The team had important support from change agents from the Welfare Administration's IT department with expertise in agile methods and a high level of trust from the IT leadership. Second, stakeholders and project leadership were given an arena in which to observe and learn how the agile approach played out in practice, instilling confidence that it was possible to make the change in the organization and domain successfully.

Running a pilot team was not without challenges. The project had to handle a "multi-speed setup" with different cadences and levels of documentation, reporting, and release routines. It made the planning of effort and coordination of the work of the teams harder, as the pilot team followed a much shorter release cycle and had different goals and priorities. The pilot team was shielded from the traditional process requirements implemented for the rest of the project, so it could focus on delivering working code and value to end-users. This situation created some tension among the remaining teams, who had to meet more stringent process requirements, such as the production of extensive documentation, detailed estimations, and detailing of solution descriptions for user stories.

**Ensure technical preparations are in place to support the transition**
Before the project changed to continuous delivery, every release had to be coordinated across the organization in a release train every three to six months. After the transition to the new delivery model, the project made at least daily deliveries to production. To achieve the change to continuous delivery, it was essential to set up a new technical self-service platform. This platform gave the teams flexibility in creating and maintaining services and established the foundation for frequent deliveries without barriers in process or organization.

Having ten development teams working on the same codebase caused a bottleneck. In the first phase of the project, the produced solution was based on further development of a prototype following a "monolith first" approach [14]. The product was built for automatic and semi-automatic handling of parental leave cases in a domain created by politics, negotiations, and law over decades. Boundaries between functional domains, known as "bounded contexts" in domain-driven design, had to be established, based on experience, by building the system in close collaboration with senior caseworkers. New technical modules were established as functional and technical dependencies stabilized and the case management domain was further understood. The transition focused on breaking up the product into more cohesive modules and decreasing coupling across identified boundaries. Modules with many dependencies, which gathered information from many registers or messages from users, were separated into new microservices.

Another bottleneck was related to the flow of code from commit to production. In the first phase of the project, teams made changes in branches of the codebase, using a Gitflow [15] branching model, which were later integrated and released to various test environments in a three-week iteration. Many teams had week-long branches, which gave rise to conflicts when merging code. Small, low-risk changes would take up to half an hour to integrate across the build stages. In the transition, the project moved to trunk-based development [16], where individual code contributions were integrated at least daily, reducing the size of merges and making work in progress more transparent to other developers.

A team of seniors with prior experience in the platform and with continuous deployment led technical fora and meetings, including fora for "tech review," a "tech lead forum," meetings on "go/no-go" for production, and "blameless postmortems." A new test regime was built on the agile test pyramid [17], and much effort was invested in removing slow and unstable tests from the main build to deliver on quality post-transition.

**Involve the project organization early and often in the transition process**
Early involvement in the transition process was ensured by establishing a task force long before the decision to change the delivery model was taken. The aim of establishing the task force was to create, at an early stage, a common understanding of the goal of the transition, the delivery model to use, and what it would require to succeed. The task force was asked to agree on the direction of the change, present a united goal, and create an initial plan to achieve the goal. The task force was cross-functional with, in total, 12 representatives from development, architecture, requirements engineering, quality, and management roles from both the Welfare Administration and external vendors participating in the project. It received input on wishes, needs, and worries before and during the implementation of the change. There were discussions on the uncertainty of remodeling the entire development organization of almost 130 people, changing roles and responsibilities, while being committed to external milestones and value realization. The task force concluded with a unified recommendation on moving to continuous deployment and autonomous teams.

The task force believed that early and extensive involvement would result in a high degree of willingness to change and to deal with risks and unknowns. Hence, it focused on ensuring everyone felt involved, creating excitement, and making those involved see opportunities rather than challenges.

We observed that early and extensive involvement from both the Welfare Administration and suppliers was key in both securing ownership throughout the organization and setting a common goal across previously strong borders between customers and vendors. Internal involvement at the Welfare Administration also generated these outcomes, with both functional and technical departments cheering the change to a more agile delivery model.

**Give autonomous teams sufficient time to become productive**
In the first phase, business people were organized in teams which were separate from the development teams, with handovers of user stories and solution descriptions before each sprint. This model was seen as a threat to the overall success of the project in the last phase, where the user needs had been less analyzed and new features had to be implemented into a running solution, increasing the need for close cooperation and communication. To address this threat, the project changed to cross-functional autonomous teams which were set up for product domains [18].

Initially, it was a challenge for which some team members were less prepared than others. When using the delivery model of the first phase, the people working on the business side had more time to prepare for the change than the developers. After the transition, the people responsible for development were on a tight schedule to meet milestones from the first phase, meaning that the teams needed time to familiarize themselves with the new model.

In order not to lose momentum in the transition, the new teams were composed of the existing Scrum-teams from the development side and additional people from the business side. In several teams, this created friction when the old internal culture persisted, only now with new "outsiders" from the business side. For some teams, it took time to develop a new common team culture with mutual trust extending across team members. As one team member put it, "*we need to have a funeral beer and mourn our old team before we can celebrate the beginning of a new team.*" However, after some weeks, most project participants perceived the cross-functional autonomous teams to outperform the previous organization.

With increased autonomy, the teams were free to decide on the design of the work processes, leading to a period characterized by lack of coordination until the teams re-installed some of the previous coordination mechanisms or developed new ones. After a while, mechanisms were clearly defined by the needs of the teams and owned by the teams collectively.

It might seem inefficient to move to autonomous teams instead of adopting a framework such as SAFe or LeSS [4, 5]. However, the teams had solid domain and technical knowledge, and it was important to focus on delivering the product and limit time spent on training or evaluating practices prescribed by all-inclusive frameworks.

The transition work reduced the cognitive load on the teams by focusing clearly on a product domain and reducing the technical complexity of the new platform [18]. Continuous deployment made the whole team work on the same set of user stories, where it could previously have taken weeks or months from when the business people were working to analyze a user story to the developers starting coding. Moving important dialogue on requirements from coordination between teams to coordination within a team was a large advantage in terms of efficiency and effectiveness.

**Move the focus from technical deliverables to continuous work on delivering user benefits**
The planned realization of the project's user benefits was ambitious and included substantial time and cost savings for the Welfare Administration and external users. Planned benefits included reduced effort to process cases with automation, increased quality of data through more self-service for end-users, and reducing the number of calls for guidance, errors in payments, and complaints. To ensure that the planned benefits would be realized, the organization introduced several benefits management roles and processes [19]. These were present from the initiation of the project, but the transition improved the ability to realize the planned benefits.

The national governmental software process recommendation is that the benefits responsible should be a manager in the "line organization." For this project, however, benefits management roles were held by people who were integrated in and very active parts of the project. This was, in particular, the case for the change manager role, but the project manager was also responsible to a considerable degree for ensuring that the planned benefits were realized. The project was able to draw on a good mix of people with responsibility, mandate, and time to act with regard to benefits management. This mix supported the project's ability to successfully realize the planned benefits.

The transition to the new delivery model enabled continuous deliveries, which were used to gain feedback and learn how well such deliveries enabled the realization of the planned benefits. Hence, there appears to be a good fit between the chosen delivery model and good benefits management.

A "we build it, we run it" regime was implemented, implying that every team was day-to-day responsible for the products in use. Bugs, requested changes from external or internal users, and planned user stories were prioritized and treated in the same way. As part of this feedback and learning process, the benefits responsible regularly communicated and followed up the planned benefits with the development teams and other stakeholders. An "impact map" visualized the connection between deliveries, impact created, and strategic goals, acting as a constant reminder that the goal of the project was not deliverables, but positive effects for the stakeholders in terms of cost and time savings and improved service levels.

The use of agile processes as part of benefits management may not only have enabled the project to prioritize deliveries according to their benefits, but also to identify new benefits and change the importance of previously planned ones. An example was extending the scope of the project to redesign external web pages, which increased the volume of digital processing of applications.

**Key learnings from the transition**

Changing the delivery model to a more agile model is a risky process. It is even more risky when it is carried out during the execution of a large software development project of high importance and when time is constrained by a fixed completion date in order for new welfare legislation to be implemented. Although the general advice is to reduce the size of software projects, and some have predicted the "death of big software," [20] it often does not make sense to modernize only a part of a legacy system. The project reported in this article managed to be successful in both changing the process and delivering a product on time. Lessons learned from this transition include the need for the careful composition of a pilot team, planning for sufficient effort for technical preparations, early and frequent involvement of stakeholders, giving autonomous teams sufficient time to establish productive work practices, and focusing on delivering benefits.

**References**
[1] Mithas, S., Kude, T., and Reisman, S., "Digitization and Disciplined Autonomy," *IT Professional*, vol. 19, pp. 4-10, 2017.
[2] Rigby, D. K., Sutherland, J., and Noble, A., "Agile at scale," Harvard Business Review, vol. 96, pp. 88-96, 2018.
[3] Conboy, K. and Carroll, N., "Implementing Large-Scale Agile Frameworks: Challenges and Recommendations," *IEEE Software,* vol. 36, pp. 44-50, 2019.
[4] Edison, H., Wang, X., and K., C., "Comparing Methods for Large-Scale Agile Software Development: A Systematic Literature Review," *IEEE Transactions on Software Engineering,* pp. 1-1, 2021.
[5] Dingsøyr, T., Falessi, D., and Power, K., "Agile Development at Scale: The Next Frontier," *IEEE Software,* vol. 36, pp. 30-38, 2019.
[6] Paasivaara, M., Behm, B., Lassenius, C., and Hallikainen, M., "Large-scale agile transformation at Ericsson: a case study," *Empirical Software Engineering,* vol. 23, pp. 2550-2596.

[7] Dikert, K., Paasivaara, M., and Lassenius, C., "Challenges and success factors for large-scale agile transformations: A systematic literature review," *Journal of Systems and Software,* vol. 119, pp. 87-108, 2016/09/01/ 2016.

[8] By, R. T., "Organisational change management: A critical review," *Journal of Change Management*, vol. 5, pp. 369-380, 2005/12/01 2005. 10.1080/14697010500359250

[9] Dingsøyr, T., Bjørnson, F. O., Schrof, J., and Sporsem, T., "A longitudinal explanatory case study of coordination in a very large development programme: The impact of transitioning from a first- to a second-generation large-scale agile development method," *Empirical Software engineering*, 2023.

[10] Dingsøyr, T., Dybå, T., Gjertsen, M., Jacobsen, A. O., Mathisen, T.-E., Nordfjord, J. O., Røe, K., and Strand, K., "Key Lessons from Tailoring Agile Methods for Large-Scale Software Development," *IEEE IT Professional,* vol. 21, pp. 34-41, 2019.

[11] Mohagheghi, P. and Lassenius, C., "Organizational implications of agile adoption: a case study from the public sector," in *ESEC/FSE*, 2021, pp. 1444-1454.

[12] Fitzgerald, B. and Stol, K.-J., "Continuous software engineering: A roadmap and agenda," *Journal of Systems and Software*, vol. 123, pp. 176-189, 2017.

[13] Ahmad, M. O., Markkula, J., and Oivo, M., "Kanban in software development: A systematic literature review," in 2013, *Euromicro,* 2013, pp. 9-16.

[14] Fowler, M. (2015). *MonolithFirst*. https://martinfowler.com/bliki/MonolithFirst.html.

[15] Driessen, V. (2010). A successful Git branching model. https://nvie.com/posts/a-successful-git-branching-model/

[16] Hammant, P. (2017). *Trunk-based development: Introduction*. https://trunkbaseddevelopment.com/

[17] Fowler, M. (2012). *Test pyramid*. http://martinfowler.com/bliki/TestPyramid.html

[18] Skelton, M. and Pais, M., *Team Topologies: Organizing Business and Technology Teams for Fast Flow*: It Revolution, 2019.

[19] Jørgensen, M., "A survey on the characteristics of projects with success in delivering client benefits," *Information and Software Technology,* vol. 78, pp. 83-94, 2016/10/01/ 2016.

[20] Andriole, S. J., "The death of big software," *Communications of the ACM,* vol. 60, pp. 29-32, 2017.

**Author bios**

Torgeir Dingsøyr is Professor in Software Engineering – Agile at the Department of Computer Science, Norwegian University of Science and Technology. He is also Adjunct Chief Research Scientist at the SimulaMet Research Laboratory. His research has focused on teamwork and learning in software development, as well as development methods for large software projects and programs. He has published in the software engineering, information systems, and project management fields. He can be contacted at Torgeir.dingsoyr@ntnu.no.

Magne Jørgensen received a Dr. Scient degree in Informatics from University of Oslo in 1994. He has experience as a software developer, project leader, and manager. He is currently a Chief Research Scientist at Simula Metropolitan and a professor at Oslo Metropolitan University and University of Oslo. Current research interests are software management, agile development, psychology of human judgement, and cost estimation. He can be contacted at magnej@simula.no

Frode Odde Carlsen is an architect, tech lead, and developer in Sopra Steria. He was the principal solution architect designing case management solutions for most of the first two releases and part of the technical leadership team post-transition. He is a long-term proponent of agile for large-scale development and part of the task force to change the project operating model to continuous delivery. He can be contacted at frode.carlsen@soprasteria.com

Lena Carlström is Managing Director, Health and Public Service, at Accenture. She was assisting project director of the Parental Benefit Project, and has an MSc in industrial engineering and management from Linköping University, Sweden (2006).

Jens Engelsrud is Head of Advisory, Core Application, at Sopra Steria. He led the solutions team in the development project. He has a Master's in Political Science and Government from the University of Oslo.

Kine Hansvold is a manager in Accenture with a background in engineering and mathematics. She was part of the Parental Benefit Project from 2017 until the project was handed over to operations and worked in one of the autonomous teams after the transition. Since the project ended, Kine has continued working with agile projects in the public sector.

Mari Heibø-Bagheri is an agile coach, service designer, and head of the "experience agency" Accenture Interactive in Norway. She was responsible for planning the self-service pilot and team lead for the autonomous, cross-functional, self-service team. Mari has worked for several award-winning programs in the public sector and has more than 14 years' experience in working with agile transformation, design, and innovation for larger organizations with extensive application portfolios. She can be contacted at mari.heibo@accenture.com

Kjetil Røe is a Consulting Director at Sopra Steria, where he has worked on several large-scale agile projects. He was project manager for the development project in the first phase of the Parental Benefit Project. He is PMP-certified and received an MSc in Engineering Cybernetics from the Norwegian University of Science and Technology, Trondheim, in 1992. He can be contacted at kjetil.roe@soprasteria.com.

Karl Ove Vika Sørensen is Project Director at Sopra Steria. He was responsible for construction in the development project. He received an MSc in Computer Science from the Norwegian University of Science and Technology, Trondheim, in 1994.